\documentclass[sigconf]{acmart}
\usepackage[utf8]{inputenc}
\usepackage[T1]{fontenc}
\usepackage{hyphenat}
\usepackage{xspace}
\usepackage{amsmath}
\usepackage{amsfonts}
\usepackage{hyperref}
\usepackage{url}
\usepackage{booktabs}
\usepackage{multirow}
\usepackage{subfigure}
\usepackage{makecell}
\usepackage{caption}
\usepackage{minibox}
\usepackage{bbm}
\usepackage{graphicx}
\usepackage{balance}
\usepackage{mathtools}
\usepackage{color}
\usepackage{marvosym}
\usepackage{ifthen}
\usepackage{textcomp}
\usepackage{enumitem}
\usepackage{verbatim}
\usepackage{algorithm}
\usepackage{algorithmic}
\usepackage{numprint}

\usepackage{amsthm}
\theoremstyle{plain}

\definecolor{MyRed}{rgb}{0.6,0.0,0.0} 
\definecolor{MyBlack}{rgb}{0.1,0.1,0.1} 
\newcommand{\inred}[1]{{\color{MyRed}\sf\textbf{\textsc{#1}}}}

\newcommand{\frameit}[2]{
  \begin{center}
  {\color{MyRed}
  \framebox[.9\columnwidth][l]{
    \begin{minipage}{.85\columnwidth}
    \inred{#1}: {\sf\color{MyBlack}#2}
    \end{minipage}
  }\\
  }
  \end{center}
}

\newcommand{\chatoDisplayMode}[1]{#1}

\newcommand{\note}[2][]{\chatoDisplayMode{\def\@tmpsig{#1}\frameit{{\Pointinghand} Note}{#2\ifx \@tmpsig \@empty \else \mbox{ --\em #1}\fi}}}
\newcommand{\todo}[2][]{\chatoDisplayMode{\def\@tmpsig{#1}\frameit{{\Writinghand} To-do}{#2\ifx \@tmpsig \@empty \else \mbox{ --\em #1}\fi}}}

\newcommand{\abbrevStyle}[1]{#1}

\newcommand{\ie}{\abbrevStyle{i.e.}\xspace}
\newcommand{\eg}{\abbrevStyle{e.g.}\xspace}
\newcommand{\cf}{\abbrevStyle{cf.}\xspace}

\newcommand{\etc}{\abbrevStyle{etc.}\xspace}

\newcommand{\Secref}[1]{Sec.~\ref{#1}}
\newcommand{\Eqnref}[1]{(\ref{#1})}

\newcommand{\Figref}[1]{Fig.~\ref{#1}}

\newcommand{\Appref}[1]{Appendix~\ref{#1}}

\newcommand{\xhdr}[1]{\vspace{1.7mm}\noindent{{\bf #1.}}}

\newcommand{\textcite}[1]{\citeauthor{#1} \shortcite{#1}}

\newcommand{\cpt}[1]{\textsc{\MakeLowercase{#1}}}

\newcommand{\hide}[1]{}

\makeatletter
\newcommand{\iffont}[2]{\ifthenelse{\equal{\f@family}{#1}}{#2}{}}
\makeatother

\iffont{ptm}{
  \usepackage{mathptmx}

  \DeclareSymbolFont{greek}{OML}{cmm}{m}{n}
  \DeclareMathSymbol{\alpha}{\mathalpha}{greek}{"0B}
  \DeclareMathSymbol{\beta}{\mathalpha}{greek}{"0C}
  \DeclareMathSymbol{\gamma}{\mathalpha}{greek}{"0D}
  \DeclareMathSymbol{\delta}{\mathalpha}{greek}{"0E}
  \DeclareMathSymbol{\epsilon}{\mathalpha}{greek}{"0F}
  \DeclareMathSymbol{\zeta}{\mathalpha}{greek}{"10}
  \DeclareMathSymbol{\eta}{\mathalpha}{greek}{"11}
  \DeclareMathSymbol{\theta}{\mathalpha}{greek}{"12}
  \DeclareMathSymbol{\iota}{\mathalpha}{greek}{"13}
  \DeclareMathSymbol{\kappa}{\mathalpha}{greek}{"14}
  \DeclareMathSymbol{\lambda}{\mathalpha}{greek}{"15}
  \DeclareMathSymbol{\mu}{\mathalpha}{greek}{"16}
  \DeclareMathSymbol{\nu}{\mathalpha}{greek}{"17}
  \DeclareMathSymbol{\xi}{\mathalpha}{greek}{"18}
  \DeclareMathSymbol{\pi}{\mathalpha}{greek}{"19}
  \DeclareMathSymbol{\rho}{\mathalpha}{greek}{"1A}
  \DeclareMathSymbol{\sigma}{\mathalpha}{greek}{"1B}
  \DeclareMathSymbol{\tau}{\mathalpha}{greek}{"1C}
  \DeclareMathSymbol{\upsilon}{\mathalpha}{greek}{"1D}
  \DeclareMathSymbol{\phi}{\mathalpha}{greek}{"1E}
  \DeclareMathSymbol{\chi}{\mathalpha}{greek}{"1F}
  \DeclareMathSymbol{\psi}{\mathalpha}{greek}{"20}
  \DeclareMathSymbol{\omega}{\mathalpha}{greek}{"21}
  \DeclareMathSymbol{\varepsilon}{\mathalpha}{greek}{"22}
  \DeclareMathSymbol{\vartheta}{\mathalpha}{greek}{"23}
  \DeclareMathSymbol{\varpi}{\mathalpha}{greek}{"24}
  \DeclareMathSymbol{\varrho}{\mathalpha}{greek}{"25}
  \DeclareMathSymbol{\varsigma}{\mathalpha}{greek}{"26}
  \DeclareMathSymbol{\varphi}{\mathalpha}{greek}{"27}
  \DeclareSymbolFont{otone}{OT1}{cmr}{m}{n}
  \DeclareMathSymbol{\Gamma}{\mathalpha}{otone}{0}
  \DeclareMathSymbol{\Delta}{\mathalpha}{otone}{1}
  \DeclareMathSymbol{\Theta}{\mathalpha}{otone}{2}
  \DeclareMathSymbol{\Lambda}{\mathalpha}{otone}{3}
  \DeclareMathSymbol{\Xi}{\mathalpha}{otone}{4}
  \DeclareMathSymbol{\Pi}{\mathalpha}{otone}{5}
  \DeclareMathSymbol{\Sigma}{\mathalpha}{otone}{6}
  \DeclareMathSymbol{\Upsilon}{\mathalpha}{otone}{7}
  \DeclareMathSymbol{\Phi}{\mathalpha}{otone}{8}
  \DeclareMathSymbol{\Psi}{\mathalpha}{otone}{9}
  \DeclareMathSymbol{\Omega}{\mathalpha}{otone}{10}
  \DeclareSymbolFont{syms}{OML}{cmm}{m}{it}
  \DeclareMathSymbol{\partial}{\mathord}{syms}{"40}
  \DeclareMathAlphabet{\mathbold}{OML}{cmm}{b}{it}
  \DeclareSymbolFont{largesymbols}{OMX}{cmex}{m}{n}

}

\newcommand{\GitHubURL}{\url{https://github.com/epfl-dlab/GoogleTrendsAnchorBank}}
\newcommand{\GT}{Google Trends\xspace}
\newcommand{\refq}{Q}
\newcommand{\fullname}{Google Trends Anchor Bank\xspace}
\newcommand{\shortname}{G-TAB\xspace}

\copyrightyear{2020}
\acmYear{2020}
\setcopyright{acmlicensed}\acmConference[CIKM '20]{Proceedings of the 29th ACM International Conference on Information and Knowledge Management}{October 19--23, 2020}{Virtual Event, Ireland}
\acmBooktitle{Proceedings of the 29th ACM International Conference on Information and Knowledge Management (CIKM '20), October 19--23, 2020, Virtual Event, Ireland}
\acmPrice{15.00}
\acmDOI{10.1145/3340531.3412075}
\acmISBN{978-1-4503-6859-9/20/10}

\begin{document}
\fancyhead{}

\title{Calibration of Google Trends Time Series}

\author{Robert West}
\affiliation{EPFL}
\email{robert.west@epfl.ch}



\begin{abstract}
Google Trends is a tool that allows researchers to analyze the popularity of Google search queries across time and space. In a single request, users can obtain time series for up to 5 queries on a common scale, normalized to the range from 0 to 100 and rounded to integer precision. Despite the overall value of Google Trends, rounding causes major problems, to the extent that entirely uninformative, all-zero time series may be returned for unpopular queries when requested together with more popular queries. We address this issue by proposing \textit{Google Trends Anchor Bank (G-TAB),} an efficient solution for the calibration of Google Trends data. Our method expresses the popularity of an arbitrary number of queries on a common scale without being compromised by rounding errors. The method proceeds in two phases. In the offline preprocessing phase, an ``anchor bank'' is constructed, a set of queries spanning the full spectrum of popularity, all calibrated against a common reference query by carefully chaining together multiple Google Trends requests. In the online deployment phase, any given search query is calibrated by performing an efficient binary search in the anchor bank. Each search step requires one Google Trends request, but few steps suffice, as we demonstrate in an empirical evaluation. We make our code publicly available as an easy-to-use library at \GitHubURL.
\end{abstract}

\maketitle

{\fontsize{8pt}{8pt} \selectfont
\textbf{ACM Reference Format:}\\
Robert West.
2020.
 Calibration of Google Trends Time Series.
In
\textit{Proceedings of the 29th ACM International Conference on Information and Knowledge Management (CIKM '20), October 19--23, 2020, Virtual Event, Ireland.}
ACM, New York, NY, USA, 4 pages. \url{https://doi.org/10.1145/3340531.3412075}
}

\section{Introduction}
\label{sec:Introduction}


Google makes aggregate statistics about the popularity of search queries publicly available via \GT.
The platform has become an important tool for researchers across disciplines
(\eg,
health \cite{ginsberg2009detecting} or
economics \cite{choi2012predicting}),
journalists \cite{stephens-davidowitz2012racist},
and many others.

\GT is accessible via an official Web interface,%
\footnote{\url{https://www.google.com/trends}}
as well as via unofficial APIs wrapping the Web interface.%
\footnote{
\url{https://github.com/GeneralMills/pytrends},
\url{https://github.com/PMassicotte/gtrendsR}
}
Users of \GT specify as input
up to 5 search queries (or ``topics''),
a time range,
and a geographic region.
Search queries may be specified as plain text (\eg, ``Munich population'')
or as entity identifiers from the Freebase knowledge base \cite{bollacker2008freebase} (\eg, /m/02h6\_6p for \cpt{Munich}).
Freebase identifiers are particularly handy as they allow for grouping various surface forms relating to the same topic, across languages (the English queries ``Munich'' and ``Munich population'', German ``M\"unchen'', Italian ``Monaco di Baviera'', \etc, are all counted toward the entity /m/02h6\_6p).

As output, \GT returns, among other things, time series of \textit{search interest} for the specified input queries.
Importantly, search interest is not returned in terms of absolute search volume, but normalized as described in the \GT FAQ:%
\footnote{\url{https://support.google.com/trends/answer/4365533?hl=en}}
``Each data point is divided by the total searches of the geography and time range it represents to compare relative popularity.
[\dots]
The resulting numbers are then scaled on a range of 0 to 100 based on a topic's proportion to all searches on all topics.''
Finally, the resulting search interest numbers are rounded to integer precision.

Whereas normalizing by geography and time is certainly useful,
scaling and rounding all numbers to integers between 0 and 100 poses considerable problems.
In particular, it is impossible to

\begin{enumerate}
    \item directly compare more than 5 search queries, even for a fixed geographic region and time span (due to scaling),
    \item directly compare queries with vastly different search interests (due to rounding errors).
\end{enumerate}

\begin{figure*}
    \centering
    \subfigure[Raw, linear scale]{
        \includegraphics[width=0.23\linewidth]{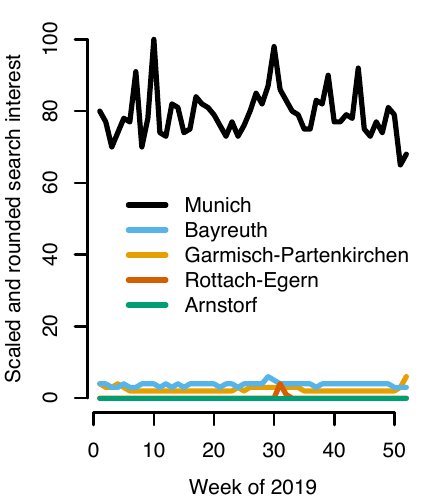}
        \label{fig:bavaria_raw_linear}
    }
    \subfigure[Raw, logarithmic scale]{
        \includegraphics[width=0.23\linewidth]{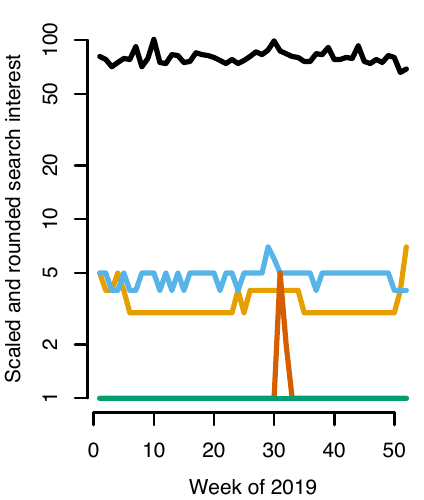}
        \label{fig:bavaria_raw_log}
    }
    \subfigure[Calibrated, logarithmic scale]{
        \includegraphics[width=0.23\linewidth]{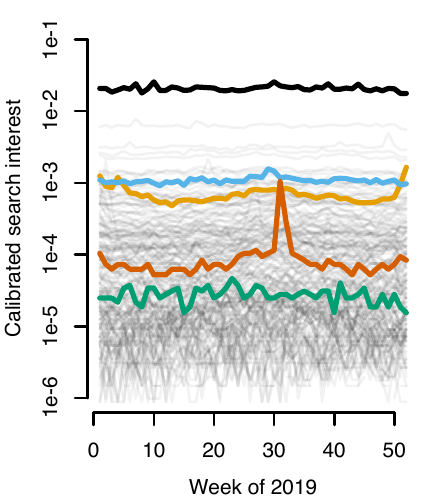}
        \label{fig:bavaria_calib}
    }
    \subfigure[Cost of binary search]{
        \includegraphics[width=0.23\linewidth]{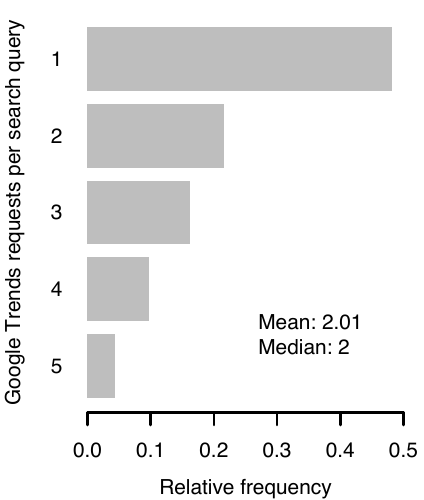}
        \label{fig:search_steps_bavaria}
    }
    \vspace{-3mm}
    \caption{
    Google search interest in Bavarian towns.
    (a)~Raw time series (scaled, rounded) for 5 towns as returned directly by \GT.
    (b)~\textit{Idem} on logarithmic scale (incremented by 1 in order to visualize values of 0).
    (c)~Calibrated time series for 200 towns as returned by our method, \shortname.
    (d)~Number of \GT requests required for calibrating one search query.
    }
    \label{fig:bavaria}
\end{figure*}

To get around problem~1, one might be tempted to always include among the up to 5 input queries a fixed reference query, against which all other queries could be compared.
But this solution will unfortunately fail for most queries due to problem~2.
\Figref{fig:bavaria}(a--b) illustrates with an example.
We used \GT to obtain the worldwide search interest in 5 Bavarian towns over the course of 2019.
The fact that the search interest in \cpt{Munich} vastly outweighs that in the 4 other towns makes the time series for the less popular towns close to useless due to rounding errors.
In particular, the time series for the least popular town, \cpt{Arnstorf}, is 0 everywhere.

We resolve these issues with
\textit{\fullname (\shortname),}
a method for obtaining calibrated \GT time series that
\begin{enumerate}
    \item are directly comparable across any number of queries (addressing problem~1) and
    \item are not compromised by rounding errors (addressing problem~2).
\end{enumerate}
 
In the above example, \shortname returns the calibrated time series of \Figref{fig:bavaria_calib}, depicted here together with the time series for about 200 more Bavarian towns.
As opposed to the raw, rounded \GT data shown in \Figref{fig:bavaria}(a--b), \shortname returns all time series with high precision across orders of magnitude, revealing details that remained hidden in the raw \GT results (\cf \Secref{sec:Results}).

We make our code publicly available as an easy-to-use library at
\GitHubURL.


\section{\fullname (\shortname)}
\label{sec:Method}

Our calibration method, \shortname, consists of two phases:
offline preprocessing and online deployment.
For a schematic description, see \Figref{fig:algo}.
All calibration is done for a given geographic region (worldwide or a specific country) and a given time span.

During \textbf{offline preprocessing}
(\Secref{sec:Offline}; left panel of \Figref{fig:algo}),
we find a set of queries that collectively span all relevant orders of magnitude in terms of search interest.
We refer to these queries as \textit{anchor queries,} and to their entirety as the \textit{anchor bank}.
One of the anchor queries is selected as the \textit{reference query} $\refq$, and each anchor query $x$ is calibrated against $\refq$ by calculating the \textit{calibrated maximum search interest} in $x$, \ie, the maximum search interest attained by $x$, divided by that attained by~$\refq$.

During \textbf{online deployment}
(\Secref{sec:Online}; right panel of \Figref{fig:algo}),
we are given an arbitrary query $q$ and return its \textit{calibrated search interest} time series, which expresses the search interest in $q$ as a fraction of the maximum search interest attained by the reference query~$\refq$.
To achieve this, we first perform a binary search in the anchor bank in order to find an anchor query $x$ whose search interest is on the same order of magnitude as that of $q$, such that $q$ and $x$ can be compared without harmful rounding errors.
Then, since the anchor query $x$ has already been calibrated against the reference query $\refq$ during offline preprocessing, the input query $q$, too, is readily calibrated against $\refq$.

\subsection{Offline phase: Building the anchor bank}
\label{sec:Offline}

\begin{figure}[b]
    \centering
    \includegraphics[width=\columnwidth]{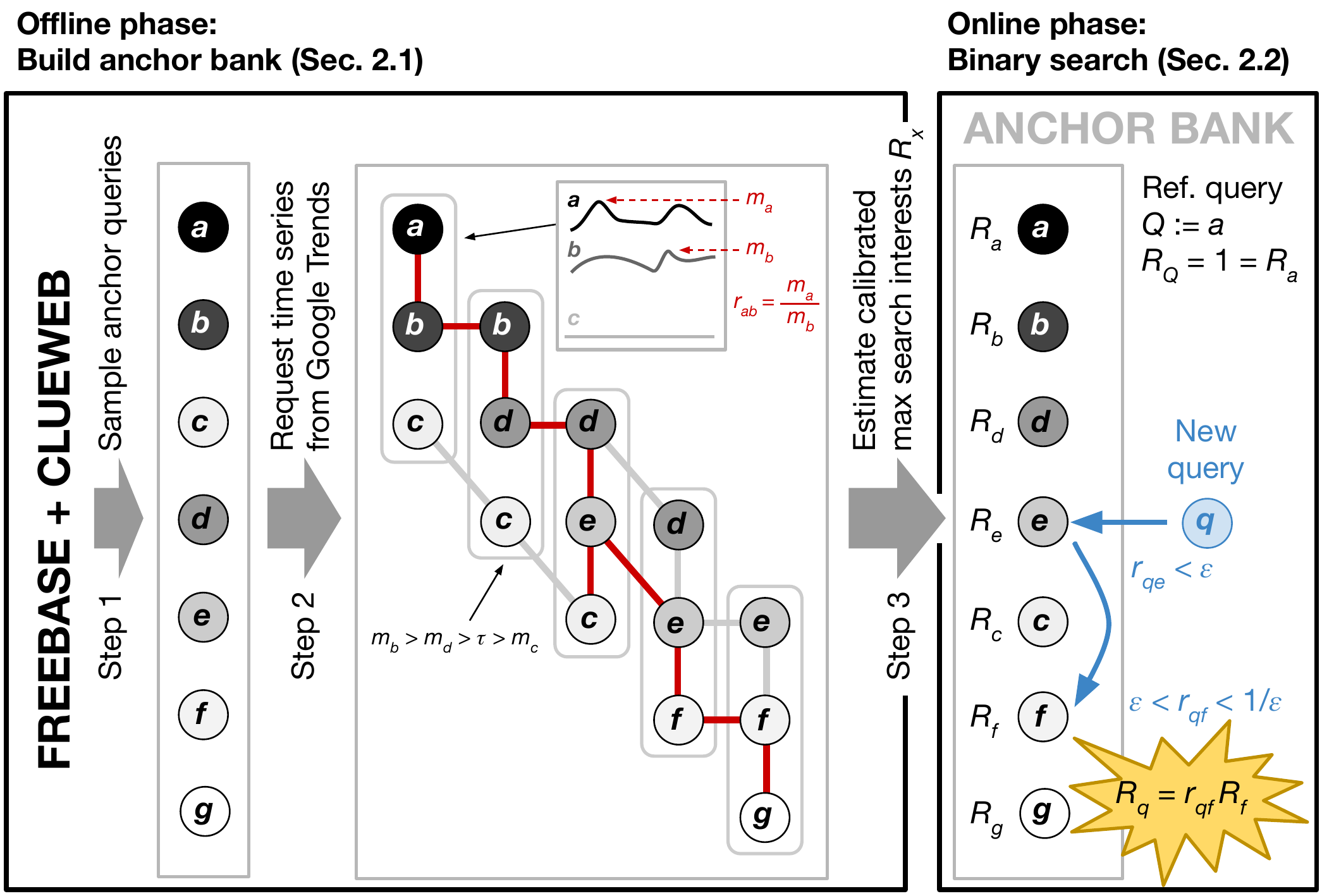}
    \caption{Overview of \fullname.}
    \label{fig:algo}
\end{figure}

In order to construct the anchor bank, the offline phase proceeds in three steps (\cf\ \Figref{fig:algo}).

\xhdr{Step 1: Sample anchor queries}
First, we identify an appropriate set of anchor queries, \ie, a set of queries that are likely to collectively cover a wide spectrum of search interest.
Anchor queries need to be found heuristically, since prior to this step, we have no information yet related to the search interest in any queries.

As a rough proxy for search interest, we use the mention frequency of Freebase~\cite{bollacker2008freebase} entities in the large-scale ClueWeb corpus~\cite{gabrilovich2013facc1}, as provided by the Freebase Easy project~\cite{bast2014easy}.
We fix an entity type with a large number of instances (\eg, persons, towns, foods),
sort the instances by ClueWeb mention frequency, discard all but the top $N$, and finally select a stratified sample of size~$n$.

As mentioned, our method calibrates all queries against a reference query $\refq$.
The reference query may be chosen manually ahead of time and added to the set of anchor queries, or it may be chosen from among the anchor queries \textit{post hoc,} \eg, as the most searched-for anchor query as determined later on, in step~3.

\xhdr{Step 2: Request time series from \GT}
Next, we sort the $n$ anchor queries by ClueWeb mention frequency and construct one \GT request for each set of $k$ contiguous queries, in a shingling fashion, for a total of $n-k+1$ requests.
In \Figref{fig:algo}, we illustrate with $k=3$ for simplicity's sake, although in practice we use the largest value allowed by \GT, $k=5$.

The purpose of constructing overlapping \GT requests is to overcome the limitations posed by rounding errors that prevent us from directly comparing queries of vastly different search interest
(\cf\ \Figref{fig:bavaria}(a--b)).
By later (in step~3) chaining together multiple overlapping \GT requests, we can compare queries across orders of magnitude in search interest transitively.

\xhdr{Step 3: Estimate calibrated maximum search interests}
For any query pair $(x,y)$, we define the \emph{maximum ratio}
\begin{equation}
    r^*_{xy} :=
    \frac{M^*_x}{M^*_y} =
    \frac{m^*_x}{m^*_y},
    \label{eqn:true max ratio}
\end{equation}
where $M^*_x, M^*_y \in \mathbb{R}_+$ denote the true, unobserved maximum search interests of $x$ and $y$ before scaling and rounding to $\{0,\dots,100\}$,
and $m^*_x, m^*_y \in [0, 100]$ denote the scaled---but still unrounded, and thus also unobserved---versions of $M^*_x$ and $M^*_y$.
The true maximum ratio $r^*_{xy}$ is unobserved, but for anchor queries $x,y$ co-occurring in the same \GT request, it can be estimated as
\begin{equation}
    r_{xy} := \frac{m_x}{m_y},
    \label{eqn:approx max ratio}
\end{equation}
where $m_x, m_y \in \{0, \dots, 100\}$ denote the maximum values attained by the scaled and rounded time series for $x$ and $y$ as returned by \GT.

The estimate $r_{xy}$ will be less reliable if $m_x$ or $m_y$ is small, due to larger relative rounding errors.
We hence discard the query pair $(x,y)$ if $\min \{m_x, m_y\} < \tau$, where $\tau \in \{0, \dots, 100\}$ is a predetermined threshold.
For instance, choosing $\tau = 10$ ensures that the two anchor queries in each retained pair fall into the same order of magnitude with respect to search interest.
One purpose of constructing \GT requests by grouping queries of similar ClueWeb mention frequency---a rough proxy for search interest---is to keep the number of discarded query pairs low.

Maximum ratios can be estimated directly only for anchor queries $x,y$ that co-occurred in at least one of the \GT requests made in step~2.
For other anchor query pairs, maximum ratios can be estimated inductively, as exemplified in \Figref{fig:algo} by the red tree connecting all anchor queries;
\eg, although the queries $a,d$ did not co-occur in any \GT request, their maximum ratio $r^*_{ad}$ can still be estimated because $a$ and $d$ both co-occurred with $b$.

Formally, maximum ratios can be inferred based on the relation
\begin{equation}
    r^*_{xz}
    = \frac{m^*_x}{m^*_z}
    = \frac{m^*_x}{m^*_y} \frac{m^*_y}{m^*_z}
    = r^*_{xy} r^*_{yz},
    \label{eqn:inference}
\end{equation}
which gives rise to a simple dynamic programming algorithm:
if the estimates $r_{xy}$ and $r_{yz}$ have already been computed, but not so $r_{xz}$, then set $r_{xz} := r_{xy} r_{yz}$.
Repeating this procedure iteratively allows us to compute $r_{xz}$ for all anchor query pairs $(x,z)$, even if $x$ and $z$ did not co-occur in any of the \GT requests.

There are in general multiple chains for transitively relating two queries $x,z$ via this procedure,
and not all chains lead to equally tight estimates $r_{xz}$ of $r^*_{xz}$:
due to integer rounding,
\begin{equation}
m^*_x \in \left[ m^-_x, m^+_x \right],
\;\;\; \text{where } \;\;\;
m^-_x = m_x-\frac{1}{2}, \;\;\; m^+_x = m_x+\frac{1}{2}
\label{eqn:m_bounds}
\end{equation}
(except if $m_x=100$, where there is no rounding, so $m^-_x=m^+_x=100$).
Hence, for $x,y$ that co\hyp occurred in the same \GT request,
\begin{equation}
r^*_{xy} \in \left[ \frac{m^-_x}{m^+_y}, \frac{m^+_x}{m^-_y} \right].
\label{eqn:r*}
\end{equation}
We can capture the tightness of the estimate $r_{xy}$ of $r^*_{xy}$ as the ratio $\eta_{xy}$ of the upper and lower bounds:
\begin{equation}
\eta_{xy} := \frac{m^+_x/m^-_y}{m^-_x/m^+_y} = \frac{m^+_x}{m^-_x} \frac{m^+_y}{m^-_y}.
\label{eqn:bound ratio}
\end{equation}
An indirect estimate $r_{xz} = r_{xy} r_{yz}$ comes with the corresponding bound ratio $\eta_{xz} = \eta_{xy} \eta_{yz}$.

For each query pair $(x,z)$, we are thus interested in finding a chain
$\langle x=x_1, x_2, \dots, x_l=z \rangle$
of queries such that any two adjacent queries $(x_i, x_{i+1})$ co\hyp occurred in the same \GT request
and the product $\eta_{xz} = \prod_{i=1}^{l-1} \eta_{x_i x_{i+1}}$ is minimized.
This problem is readily solved by finding a shortest path from $x$ to $z$ in a weighted directed graph $G$ with queries as nodes, edges between queries that co\hyp occurred in the same \GT request,
and edge weights $w_{xy} := \log \eta_{xy}$.%
\footnote{
To begin with, there are generally multiple edges per $(x,y)$, one for each \GT request in which both $x$ and $y$ occurred.
As we are interested in shortest paths in $G$, we define the edge for $(x,y)$ in $G$ based on the request with the smallest $\eta_{xy}$.
}
The product $r_{xz} = \prod_{i=1}^{l-1} r_{x_i x_{i+1}}$ along the shortest path from $x$ to $z$ yields the tightest estimate of $r^*_{xz}$ that is possible given the \GT results at hand.

The goal of offline preprocessing is to calibrate each anchor query $x$ against the reference query $\refq$, as captured by the maximum ratio $r_{x\refq}$, henceforth termed $x$'s \textit{calibrated maximum search interest}
\begin{equation}
    R_x := r_{x\refq}.
\label{eqn:R_x}
\end{equation}
The calibrated maximum search interest $R_x$ expresses the maximum search interest in query $x$ as a fraction of the maximum search interest in the reference query $\refq$.
The result of offline preprocessing---the anchor bank---consists of the list of anchor queries $x$ sorted in increasing order of their calibrated maximum search interests $R_x$.

In the next section, we describe how to efficiently calibrate \GT results for any arbitrary query~$q$, rather than for anchor queries only.


\xhdr{Obtaining an optimal anchor bank}
In
Appendix~A,%
\footnote{Appendices available online at \url{https://arxiv.org/abs/2007.13861}}
we show that, in order to obtain the most precisely calibrated maximum search interests (\ie, subject to the lowest rounding errors), the anchor bank should consist of a list of queries such that neighboring queries $x,y$ have a constant maximum ratio $r_{xy} \approx 1/e \approx 0.37$.
As shown in
Appendix~B,
this ideal can be approximated in a second round of \GT requests, based on the results of the above\hyp described first round.


\subsection{Online phase: Binary search}
\label{sec:Online}

In the online deployment phase, we are given any Google query $q$ and return a calibrated time series of search interest, \ie, a time series of $q$'s search interest expressed as a fraction of the maximum search interest achieved by the reference query $\refq$ (in the specified geographic region and time span).
This is achieved by multiplying $q$'s uncalibrated search interest time series, as obtained directly from \GT, by $R_q/m_q$.
The task, thus, is to compute $R_q$.

In principle, $R_q := r_{q\refq}$ could be measured by including $q$ and $\refq$ in the same \GT request and estimating their maximum ratio $r_{q\refq} = m_q/m_\refq$.
In practice, however, this direct estimate may be unusable due to errors incurred by rounding $m_q$ and $m_\refq$ to integer precision.
To overcome this issue, we observe that, for any anchor query $x$, the following equality holds:
\begin{equation}
    R_q = r_{q\refq} = r_{qx} r_{x\refq} = r_{qx} R_x.
    \label{eqn:online}
\end{equation}
Here, $R_x$ is already known from offline preprocessing, and the maximum ratio $r_{qx} = m_q/m_x$ can be reliably computed from the uncalibrated \GT results,
provided that $m_q,m_x \in \{0,\dots,100\}$ are both reasonably large, for then rounding errors will be small.

The main challenge of the online phase is therefore to search the anchor bank for an anchor query $x^*$ such that neither $m_q$ nor $m_{x^*}$ is too small in the result returned by \GT for a joint request for $q$ and $x^*$.
To quickly find such an anchor query $x^*$, we apply binary search:
Let $A$ be the anchor bank, \ie, a list of all anchor queries $x$ in increasing order of $R_x$.
Compare $q$ to the mid point $x'$ of $A$.
If $\epsilon < r_{qx'} < 1/\epsilon$ (where $\epsilon < 1$ is a fixed parameter),
terminate the search with $x^*:=x'$.
Otherwise, if $r_{qx'} < \epsilon$, recursively search in the left half of $A$,
and if $r_{qx'} > 1/\epsilon$, in the right half of $A$.

Computing each maximum ratio $r_{qx'} = m_q/m_{x'}$ requires sending a joint \GT request for the two queries $q$ and $x'$.
In practice, the search terminates after a very small number of steps (\cf\ \Secref{sec:Results}), so the overhead incurred by calibration is generally low.

\xhdr{Quantifying uncertainty}
Due to integer rounding, the estimate $R_q$ is approximate.
To quantify the uncertainty, we compute upper and lower bounds for $R_q$, as well as for $q$'s calibrated time series of search interest, based on the calculations of step~3 in \Secref{sec:Offline}.


\section{Example results}
\label{sec:Results}

We now showcase the power of \shortname empirically.
In step~1 of the offline phase (\Secref{sec:Offline}), we consider as anchor queries all food entities from Freebase (types \cpt{Food} and \cpt{Dish}) and sample $n=100$ from the top $N=2000$ entitites in a stratified manner.

Search interest generally follows a heavy-tailed distribution, headed by navigational queries for common websites
\cite{baeza1999modern}.
In order to also cover these head queries, we manually add 6 common navigational queries to the anchor bank. 
Manual probing of \GT revealed \cpt{Facebook} as probably the most popular Google query, which we use as the reference query $\refq$.

In step~2, we group $k=5$ queries per \GT request;
in step~3, we use $\tau = 10$;
and during the online phase, $\epsilon = 0.1$.

\xhdr{Example 1: Bavarian towns}
Resuming the example from \Secref{sec:Introduction}, we select the top 100 Bavarian towns mentioned most frequently in ClueWeb, unioned with 100 more sampled from the top 1000, and analyze their worldwide Google search interest during 2019.
Queries are represented as Freebase IDs, rather than plain text.

Our calibration method can place an arbitrary number of queries on a common scale, as exemplified in \Figref{fig:bavaria_calib}, which shows the time series for all 200 towns in one single plot.
We see that \cpt{Munich} is about 2\%, and \cpt{Arnstorf} about 0.003\% as popular as the reference query \cpt{Facebook}.
Moreover, although search interest spans 5 orders of magnitude, all 200 time series are available at high precision, revealing insights that remain hidden in the results obtained directly from \GT (\cf\ \Figref{fig:bavaria_raw_log}).
For instance, \cpt{Rottach-Egern} is consistently more popular than \cpt{Arnstorf}, and its popularity grows gradually in summer, rather than as an impulse in week~32.

What price do we need to pay for calibration?
\Figref{fig:search_steps_bavaria} answers this question by showing the distribution of the number of \GT requests required during the binary search in order to process one query.
We see that, on average, only two \GT requests are needed to calibrate one query.

\begin{figure}
    \centering
    \subfigure[Calibrated search interest]{
        \includegraphics[width=0.47\linewidth]{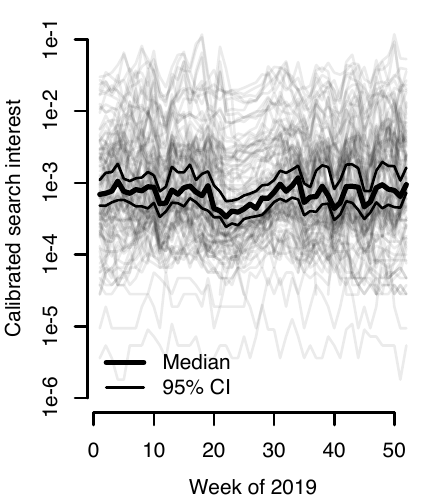}
        \label{fig:soccer_calib}
    }
    \subfigure[Cost of binary search]{
        \includegraphics[width=0.47\linewidth]{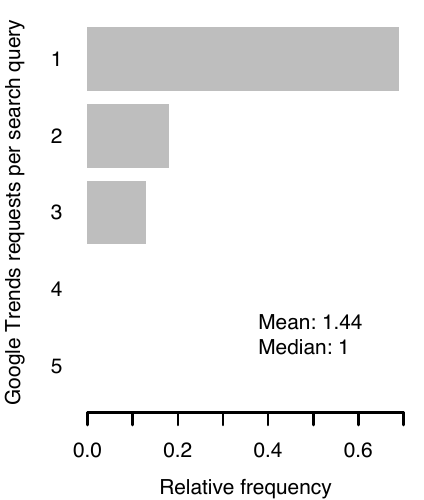}
        \label{fig:search_steps_soccer}
    }
    \vspace{-3mm}
    \caption{Search interest in top 100 soccer clubs.}
    \label{fig:soccer}
\end{figure}

\xhdr{Example 2: Soccer clubs}
As a second example, \Figref{fig:soccer_calib} plots calibrated time series for the 100 soccer clubs mentioned most frequently in ClueWeb.
The median over all clubs is shown as a thick black line (with bootstrapped 95\% confidence intervals).
It reveals that the search interest in soccer clubs drops in summer
(mid May to mid July),
when many soccer leagues break between seasons.
Without calibration, such insights could not emerge.

Calibration is even cheaper here than in example~1.
Most queries require only a single \GT request
(mean 1.44; \Figref{fig:search_steps_soccer}).
Generally, the better the distribution of $R_x$ in the anchor bank matches the distribution seen in online deployment, the more efficient the binary search.

\section{Conclusion}
\label{sec:Conclusion}

\GT has emerged as a Swiss army knife for data scientists.
So far, users have been hampered by its imprecision, stemming from the fact that all results are scaled and rounded to integers from 0 to 100.
With \shortname, our simple, effective, and efficient calibration method, we hope to have sharpened the knife for all its users.


\xhdr{Acknowledgments}
I thank Gorjan Popovski and Manoel Horta Ribeiro for help with Python coding,
and Microsoft, Google, Facebook for funding.
Dedicated to Barbara West and Johann Winkler.

\bibliographystyle{ACM-Reference-Format}

\bibliography{references}


\appendix
\section{Properties of an optimal anchor bank}
\label{sec:Characteristics of an optimal anchor bank}

An anchor bank is a list
$A=(x_1, x_2, \dots, x_n)$
of anchor queries sorted in increasing order of calibrated maximum search interest,
\ie, $R_{x_i} < R_{x_{i+1}}$.
In order to be able to compare Google queries of arbitrary search interest on the same scale using $A$, 
we assume that $x_1$ is the least frequent, and $x_n$ the most frequent, Google query (in practice, the extreme queries can be chosen heuristically by manual probing).
Without loss of generality, we assume the reference query $\refq=x_n$.
That is, in order to calibrate the anchor queries, we need to estimate $r^*_{x_i x_n}$ for all $1 \leq i < n$.

Here we ask the question:
What are the properties of an optimal anchor bank?

First, let us define optimality:
in an optimal anchor bank for fixed least and most frequent queries $x_1$ and $x_n$, the estimate $R_{x_1} = r_{x_1 x_n}$ (the estimated maximum ratio of the most distant query pair) should be as close as possible to the true, unobserved
\begin{equation}
r^* := r^*_{x_1 x_n},
\end{equation}
in the sense that the bound ratio $\eta_{x_1 x_n}$
(\cf\ \Eqnref{eqn:bound ratio})
should be as small, \ie, as close to 1, as possible.
Additionally, if multiple anchor banks meet this criterion, an anchor bank is optimal if it does so with the smallest number $n$ of anchor queries.

This definition entails that, in an optimal anchor bank $A^*$, the best way to estimate $r^*$ is via the product
\begin{equation}
r_{x_1 x_n} = \prod_{i=1}^{n-1} r_{x_i x_{i+1}},
\label{eqn:product}
\end{equation}
where each $r_{x_i x_{i+1}} = m_{x_i}/m_{x_{i+1}}$ is obtained from a pairwise \GT request $\{x_i, x_{i+1}\}$.
The reason for multiplying ratios $r_{x_i x_{i+1}}$ of neighboring queries, rather than directly comparing non\hyp neighboring queries $x_i, x_j$ (with $j>i+1$) in a separate \GT request, is that, if non\hyp neighboring queries could be compared directly without increasing the overall rounding error, then the intermediate queries $x_{i+1}, \dots, x_{j-1}$ would be unnecessary, and $A^*$ would not be optimal---a contradiction.
The reason for including only $k=2$, rather than the maximum allowed $k=5$, queries per request is that $k=2$ keeps rounding errors to a minimum, as only one of the maximum values $m_{x_i}, m_{x_{i+1}}$ will be rounded after scaling.

Next, we analyze the bound ratio $\eta_{x_1 x_n}$ that arises when constructing an anchor bank in the above\hyp described way.
Each pairwise \GT request $\{x_i, x_{i+1}\}$ returns
$m_{x_{i+1}} = 100$ and
$m_{x_i} = 100 c_i < 100$, for $c_i = r_{x_i x_{i+1}} \in [0,1]$.
Recall from \Eqnref{eqn:m_bounds} that $m_{x_i}$ is the rounded version of the true, unobserved $m_{x_i}^* \in [m_{x_i}^-, m_{x_i}^+]$ (the maximum search interest of $x_i$ after scaling to $[0,100]$, but before integer rounding), with
$m_{x_i}^- = m_{x_i}-\frac{1}{2}$, and $m_{x_i}^+ = m_{x_i}+\frac{1}{2}$.
Since $m_{x_{i+1}}=100$, we have $m_{x_{i+1}}^- = m_{x_{i+1}}^+ = 100$ (only scaling, no rounding),
so the true maximum ratio
$r^*_{x_i x_{i+1}} \in [m_{x_i}^-/100, m_{x_i}^+/100]$,
and the ratio of the upper and lower bounds is
\begin{equation}
\eta_{x_i x_{i+1}}
\;\;=\;\; \frac{m_{x_i}^+}{m_{x_i}^-}
\;\;=\;\; \frac{100c_i + \frac{1}{2}}{100c_i - \frac{1}{2}},
\;\;=\;\; \frac{c_i + \epsilon}{c_i - \epsilon},
\end{equation}
where $\epsilon=1/200$.
When multiplying local maximum ratios (\cf\ \Eqnref{eqn:product}), rounding errors accumulate, yielding the global bound ratio
\begin{equation}
\eta_{x_1 x_n} = \prod_{i=1}^{n-1} \frac{c_i + \epsilon}{c_i - \epsilon}.
\label{eqn:eta_prod}
\end{equation}

In fact, in an optimal anchor bank, anchor queries are equidistant, in the sense that $c_i=c$ for a global constant $c$.
To see this, consider the case $n=3$ (the case $n>3$ follows by induction):
\begin{equation}
\eta_{x_1 x_3}
= \left(\frac{c_1+\epsilon}{c_1-\epsilon}\right)
  \left(\frac{c_2+\epsilon}{c_2-\epsilon}\right),
\end{equation}
which is minimized for $c_1 = c_2$:
The numerator,
\begin{equation}
(c_1+\epsilon) (c_2+\epsilon)
= c_1 c_2 + \epsilon (c_1 + c_2) + \epsilon^2,
\end{equation}
is minimized for $c_1 = c_2$ because 
$c_1 c_2 = r^*$ is fixed
and the circumference $2(c_1 + c_2)$ of a rectangle of fixed area $c_1 c_2$ is minimized by a square.
Analogously, the denominator is maximized for $c_1 = c_2$, so $\eta_{x_1 x_3}$ as a whole is minimized for  $c_1 = c_2 =: c$.

The product of \Eqnref{eqn:eta_prod} thus simplifies to
\begin{equation}
\eta_{x_1 x_n}
= \left( \frac{c+\epsilon}{c-\epsilon} \right)^{n-1}
= \left( \frac{c+\epsilon}{c-\epsilon} \right)^{\log_c r^*},
\label{eqn:eta_prod_simple}
\end{equation}
and finding an optimal anchor bank boils down to finding the optimal value of the constant $c$.
Writing the bound ratio $\eta_{x_1 x_n} = \eta(c)$ as a function of $c$,
\begin{equation}
\eta(c) = \left( \frac{c+\epsilon}{c-\epsilon} \right)^{\log_c r^*},
\label{eqn:bound ratio optimal}
\end{equation}
our goal is to minimize $\eta(c)$ subject to the constraint that $r^* = c^{n-1}$ for some integer $n$, \ie, that $c$ should be a root of $r^*$.

Since the maximum ratio $r^*$ of the most and the least frequent queries is fixed, a smaller $c<1$ will lead to an anchor bank of a smaller size $n$.
This leads to a tradeoff:
for smaller $c$ (\ie, smaller $n$), we will on the one hand require fewer factors for computing $r_{x_1 x_n}$ (\cf\ \Eqnref{eqn:product}),
but on the other hand each factor will be subject to larger rounding errors (\cf\ \Eqnref{eqn:eta_prod_simple}).
Finding the optimal constant $c$ is to find the optimal tradeoff between the number of comparisons and the precision of comparisons.

\begin{figure}
    \centering
    \includegraphics[width=\linewidth]{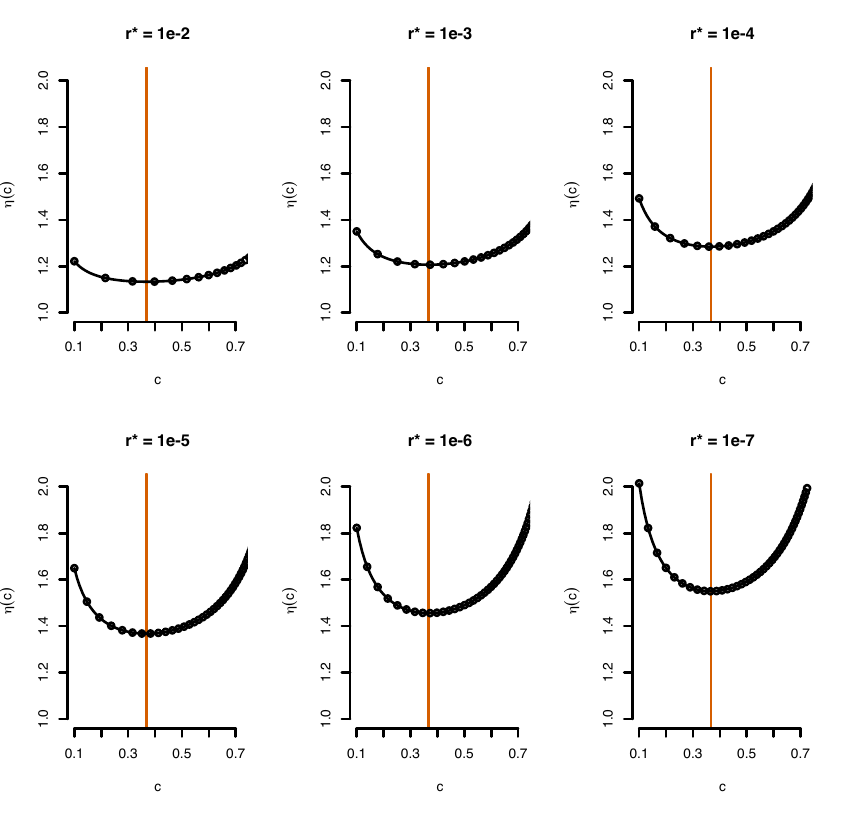}
    \caption{
    Bound ratio $\eta(c)$ achieved by an anchor bank where neighboring queries $x_i, x_{i+1}$ have a constant maximum ratio $m_{x_i}/m_{x_{i+1}} = c$, for various choices of $r^*$ (the maximum ratio of the least and most frequent queries in the anchor bank).
    \textit{Black dots:} roots of $r^*$, \ie, values of $c$ that meet the constraint $r^* = c^{n-1}$ for some integer $n$.
    \textit{Red:} value $c=1/e \approx 0.37$ for which $\eta(c)$ is approximately minimized.
    }
    \label{fig:eta_matrix}
\end{figure}

To make the problem easier (and because $r^*$, though fixed, is unobserved), we consider the continuous relaxation without the constraint $r^* = c^{n-1}$.
(We shall see later, in \Figref{fig:eta_matrix}, that adding the constraint does not change the optimal $c$ by much.)
This way, we can find the optimal $c$ by setting the derivative $\eta'(c)$ to zero:
\begin{equation}
\eta'(c)
= \eta(c) (\log r^*) \left( \frac{ \frac{1}{c+\epsilon} - \frac{1}{c-\epsilon} }{\log c} - \frac{\log \frac{c+\epsilon}{c-\epsilon}}{c (\log c)^2} \right)
= 0.
\end{equation}

Since $\eta(c) (\log r^*) < 0$, we have $\eta'(c)=0$ iff
\begin{eqnarray}
\frac{\frac{1}{c+\epsilon} - \frac{1}{c-\epsilon}}{\log c}
&=& \frac{\log \frac{c+\epsilon}{c-\epsilon}}{c (\log c)^2} \\
(\log c) \left( \frac{c}{c+\epsilon} - \frac{c}{c-\epsilon} \right)
&=& \log \frac{c+\epsilon}{c-\epsilon} \\
(\log c) \left( \frac{c}{c+\epsilon} - \frac{c}{c-\epsilon} \right)
&=& \log \left( \frac{c}{c-\epsilon} + \frac{\epsilon}{c-\epsilon} \right) \\
\log c
&=& \frac{ \log \left( \frac{1}{1-\epsilon/c} + \frac{\epsilon/c}{1-\epsilon/c} \right) } { \frac{1}{1+\epsilon/c} - \frac{1}{1-\epsilon/c} }.
\label{eqn:condition}
\end{eqnarray}

First-order Taylor expansions around $\epsilon/c = 0$ yield the following approximations:
\begin{eqnarray}
\frac{1}{1+\epsilon/c} &\approx& 1 - \epsilon/c \\
\frac{1}{1-\epsilon/c} &\approx& 1 + \epsilon/c \\
\frac{\epsilon/c}{1-\epsilon/c} &\approx& \epsilon/c,
\end{eqnarray}
so \Eqnref{eqn:condition} can be approximated as
\begin{eqnarray}
\log c &\approx& \frac{\log (1+2\epsilon/c)}{-2\epsilon/c}.
\label{eqn:condition2}
\end{eqnarray}

Another first-order Taylor expansion, around $2 \epsilon/c = 0$, yields
\begin{eqnarray}
\log (1+2\epsilon/c) &\approx& 2\epsilon/c,
\end{eqnarray}
so \Eqnref{eqn:condition2} simplifies to
\begin{eqnarray}
\log c &\approx& -1 \\
c &\approx& 1/e \;\;\;\approx\;\;\; 0.37.
\end{eqnarray}

That is, in the ideal case, we would want to construct an anchor bank in which all neighboring queries $x_i, x_{i+1}$ have the constant maximum ratio $m_{x_i}/m_{x_{i+1}} \approx 1/e$.
In \Figref{fig:eta_matrix}, we plot $\eta(c)$ for choices of $r^*$ spanning all relevant orders of magnitude.
The vertical red lines confirm visually that $\eta(c)$ is minimized for $c \approx 1/e$, regardless of $r^*$.

\begin{figure}[t]
    \centering
    \includegraphics[width=\linewidth]{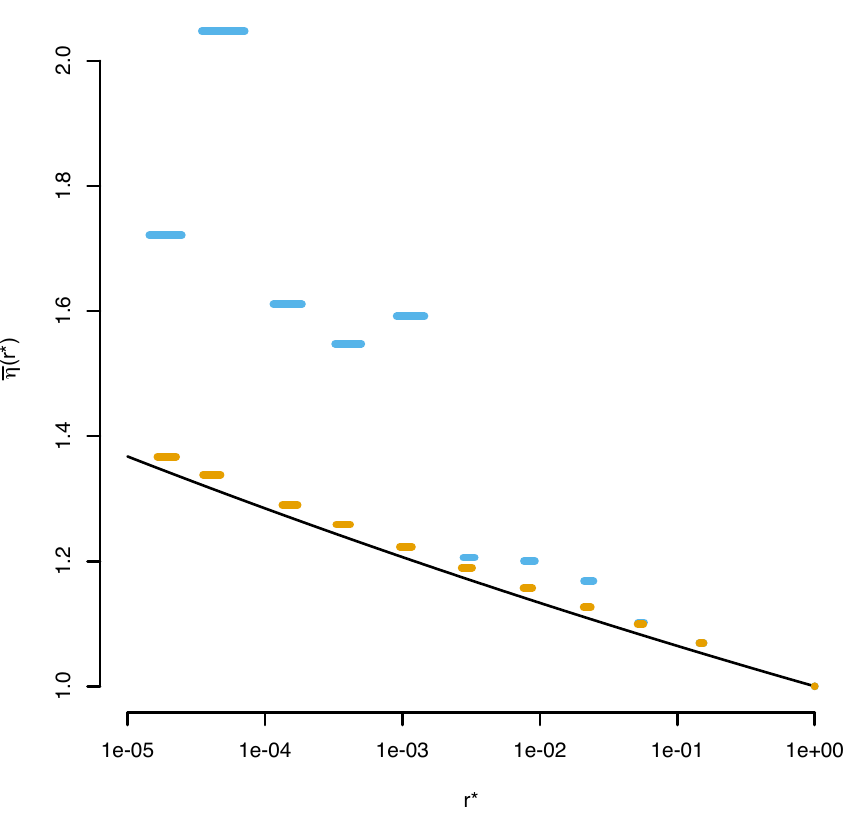}
    \caption{
    Theoretically optimal bound ratio $\bar \eta(r^*)$ (black) and empirical bound ratios (yellow, blue).
    \textit{Yellow:} empirical bound ratios $\eta_{x_i x_n}$ obtained for anchor queries $x_i$ when using the optimized anchor bank.
    \textit{Blue:} \textit{Idem,} but when using the initial anchor bank before optimization (showing only the subset of anchor queries retained for the optimized anchor bank).
    As we see, the optimized anchor bank comes much closer to the theoretical optimum (black).
    Since the exact maximum ratios $r^*_{x_i x_n}= m^*_{x_i} / m^*_{x_n}$ are unobserved, we plot, on the horizontal axis, the intervals $[ m^-_{x_i}/m^+_{x_n}, m^+_{x_i}/m^-_{x_n} ]$ instead (\cf\ \Eqnref{eqn:r*}).
    }
    \label{fig:f_comparison}
\end{figure}

We now fix $c=1/e$ and turn the bound ratio 
$\eta_{x_1 x_n}$
into a function $\bar \eta(r^*)$ of $r^*$ (rather than, as before, of $c$):
\begin{equation}
\bar \eta(r^*)
= \left( \frac{\frac{1}{e}+\epsilon}{\frac{1}{e}-\epsilon} \right)^{-\log r^*}
\approx 1.028^{-\log r^*}
\approx (r^*)^{-0.027}.
\end{equation}
In practice, $r^* > 10^{-7}$, so $\bar \eta(r^*) < 1.55$.
That is, in the worst case, the upper bound on an optimally estimated maximum ratio is at most 55\% higher than the lower bound.
We plot $\bar \eta(r^*)$ in \Figref{fig:f_comparison} as the black curve.

\section{Constructing and using an optimal anchor bank}
\label{sec:Computing an optimal anchor bank}

The considerations of \Appref{sec:Characteristics of an optimal anchor bank} where theoretical.
Next, we describe how to implement them in order to construct and deploy an approximately optimal anchor bank in practice.

\xhdr{Offline phase: Building the anchor bank}
Starting from an initial anchor bank $A$ constructed as described in \Secref{sec:Offline}, we can build an approximately optimal anchor bank $A^*$ (as characterized in \Appref{sec:Characteristics of an optimal anchor bank}) as follows:
\begin{enumerate}
\item Select a subset of the anchor queries of the initial anchor bank $A$ such that, in the subset, subsequent queries $x_i, x_{i+1}$ have an estimated maximum ratio $r_{x_i x_{i+1}}$ of about~$1/e$ (\cf\ \Appref{sec:Characteristics of an optimal anchor bank}).
\item Obtain more precise estimates of the true maximum ratios $r^*_{x_i x_{i+1}}$ of subsequent queries in a second round of \GT requests, where we include only $k=2$ (rather than $k=5$) subsequent queries $\{x_i, x_{i+1}\}$ per request.
\end{enumerate}

To implement step~1, we first construct a complete, directed graph $G'$ with edge weights
$w_{xy} := |\log(e^{-1}/r_{xy})| = |1+\log r_{xy}|$,
where $r_{xy}$ are the maximum\hyp ratio estimates computed in step~3 of offline preprocessing (\Secref{sec:Offline}),
and then find a shortest path from the least frequent query $x_1$ to the most frequent query $x_n$ in $G'$.
The nodes along the shortest paths then serve as the anchor queries of the optimized anchor bank $A^*$.

\begin{figure}[t]
    \centering
    \includegraphics[width=\linewidth]{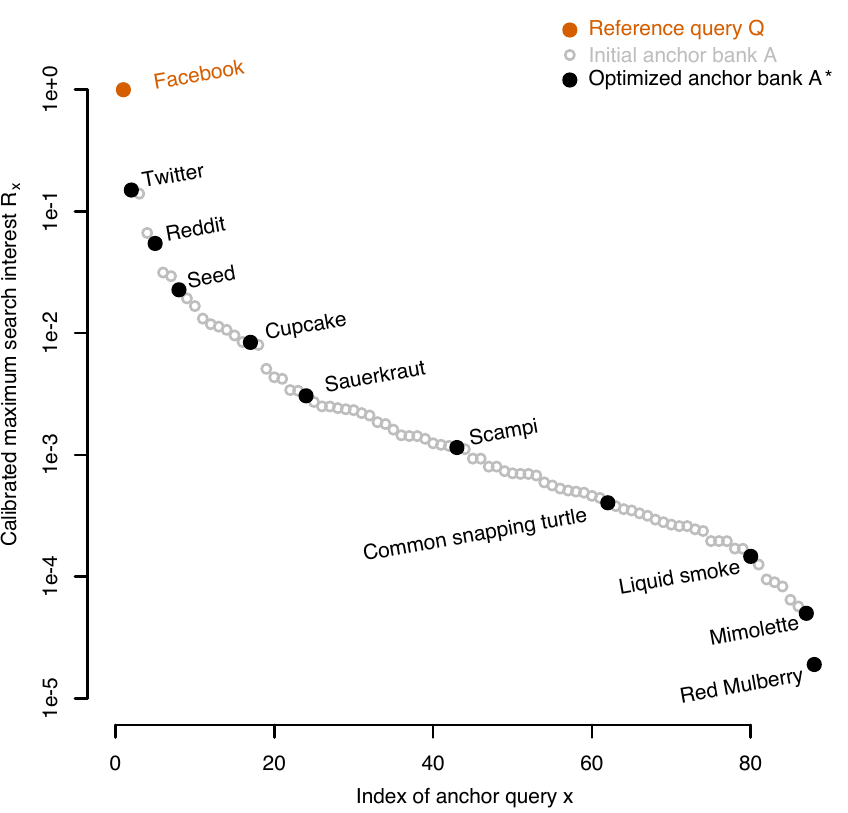}
    \caption{
    Anchor bank before and after optimization.
    \textit{Gray circles:} Initial anchor bank $A$ constructed as described in \Secref{sec:Offline}.
    \textit{Black disks:} Optimized anchor bank $A^*$.
    Note that the optimized anchor bank divides the (logarithmic) $y$-axis into segments of approximately equal length, corresponding to the optimal maximum ratio of $1/e$ for neighboring queries (\cf\ \Appref{sec:Characteristics of an optimal anchor bank}).
    }
    \label{fig:anchor_bank}
\end{figure}

In \Figref{fig:anchor_bank}, we visualize the initial anchor bank $A$ constructed as described in \Secref{sec:Offline}, as well as the optimized anchor bank $A^*$ constructed as described above.
Note that the optimized anchor bank divides the (logarithmic) $y$-axis into segments of approximately equal length, corresponding to the optimal maximum ratio of $1/e$ for neighboring queries, as derived in \Appref{sec:Characteristics of an optimal anchor bank}.

We visually demonstrate the improvements afforded by optimizing the anchor bank in \Figref{fig:f_comparison}.
Here we calibrated the anchor queries $x_i$ retained for the optimized anchor bank $A^*$ against the reference query $Q=x_n=\text{\cpt{Facebook}}$ in two ways:
first, using the initial anchor bank $A$, and second, using the optimized anchor bank $A^*$.
(Recall that the anchor queries of $A^*$ form a subset of those of $A$.)
In blue, we plot, on the vertical axis, the bound ratios $\eta_{x_i x_n}$ (\cf\ \Eqnref{eqn:bound ratio}) obtained for the initial anchor bank $A$.
In yellow, we plot the corresponding bound ratios obtained for the optimized anchor bank $A^*$.
We clearly see that the optimized anchor bank (yellow) comes much closer to the theoretically optimal bound ratio $\bar\eta(r^*_{x_i x_n})$ (black) than the initial anchor bank (blue) does.

\xhdr{Online phase: Binary search}
At deployment time, during online binary search (\Secref{sec:Online}), we start by comparing the input query $q$ to the query $x'$ from the optimized anchor bank $A^*$ that is closest (in terms of calibrated maximum search interest $R_{x'}$) to the median query of the initial anchor bank $A$, based on the following rationale:
the search\hyp interest distribution in the initial anchor bank approximates the overall search-interest distribution, so by first comparing to a query that is close to the overall median, we will reduce the number of search steps, which keeps the number of \GT requests as well as rounding errors low.

\balance

As mentioned at the end of \Secref{sec:Online}, in addition to the calibrated search\hyp interest time series, we also return upper and lower bounds of the time series based on the largest possible (unknown) rounding errors that may have been encountered.
In order to keep the bounds as tight as possible, we also use the close-to-median query $x'$ (as defined above) as the reference query $\refq$ in practice (whereas, for ease of exposition, we have used the most frequent query $x_n$ as the reference query throughout this paper).

\end{document}
\endinput